%% file: DePIN.tex
\documentclass[sigconf]{acmart}
\renewcommand\footnotetextcopyrightpermission[1]{} 

\usepackage{colortbl}
\usepackage{xcolor}
\usepackage{pifont}
\usepackage{hyperref}
\usepackage{array}
\usepackage{tikz}
\usetikzlibrary{shapes.geometric, arrows.meta, positioning, fit, backgrounds}
\usepackage{geometry}
\usepackage{array}
\usepackage{enumitem}
\usepackage{subcaption}
\AtBeginDocument{%
  }

\begin{document}

\title[EconAgentic in DePIN Markets]{EconAgentic in DePIN Markets: A Large Language Model Approach to the Sharing Economy of Decentralized Physical Infrastructure}

\author{Yulin Liu}
\authornote{Corresponding author. Email: yulin@quantecon.ai.}
\affiliation{%
  \institution{Quantum Economics}
  \city{Zug}
  \country{Switzerland}
}

\author{Mocca Schweitzer}
\affiliation{%
  \institution{Quantum Economics}
  \city{Zug}
  \country{Switzerland}
}

\renewcommand{\shortauthors}{Liu}

\begin{abstract}
The Decentralized Physical Infrastructure (DePIN) market is revolutionizing the sharing economy through token-based economics and smart contracts that govern decentralized operations. By 2024, DePIN projects have exceeded \$10 billion in market capitalization, underscoring their rapid growth. However, the unregulated nature of these markets, coupled with the autonomous deployment of AI agents in smart contracts, introduces risks such as inefficiencies and potential misalignment with human values. To address these concerns, we introduce \textit{EconAgentic}, a Large Language Model (LLM)-powered framework designed to mitigate these challenges. Our research focuses on three key areas: 1) modeling the dynamic evolution of DePIN markets, 2) evaluating stakeholders' actions and their economic impacts, and 3) analyzing macroeconomic indicators to align market outcomes with societal goals. Through \textit{EconAgentic}, we simulate how AI agents respond to token incentives, invest in infrastructure, and adapt to market conditions, comparing AI-driven decisions with human heuristic benchmarks. Our results show that \textit{EconAgentic} provides valuable insights into the efficiency, inclusion, and stability of DePIN markets, contributing to both academic understanding and practical improvements in the design and governance of decentralized, tokenized economies.
\end{abstract}





\maketitle

\section{Introduction}

The introduction of blockchain technology has been transformative across various sectors~\cite{monrat2019survey}, notably in finance, where it ensures secure and transparent financial transactions~\cite{Zhang23}, and in supply chain management, where it facilitates greater traceability and efficiency~\cite{chang2020blockchain}. Diverging from these applications, the concept of Decentralized Physical Infrastructure Networks (DePIN) emerges as a particularly innovative use of blockchain in managing physical infrastructures. DePIN employs blockchain technology to manage crucial systems such as decentralized wireless networks, cloud computing platforms, and distributed storage solutions, introducing sector-specific innovations like Helium’s network\footnote{https://www.helium.com/}, Render’s platform\footnote{https://render.com/}, and Filecoin’s storage\footnote{https://filecoin.io/} that disrupt traditional centralized models and improve infrastructure management across various sectors~\cite{cointelegraph_DePIN}. Unlike traditional systems, where management often relies on centralized hubs—such as control centers for electricity grids or operational headquarters for public transportation—which can act as bottlenecks or single points of failure, DePIN introduces a more resilient and democratic approach to infrastructure management.

As of 2024, the traction for DePIN is evident, with market capitalizations exceeding \$10 billion and even over \$30 billion in May 2024, according to \textit{CoinMarketCap}\footnote{https://coinmarketcap.com/view/DePIN/} and \textit{DePINscan}\footnote{https://DePINscan.io/}. This substantial growth reflects a significant shift toward decentralization, promising to address the inefficiencies and vulnerabilities inherent in centralized systems and aligning with the foundational principles of Web3.

The Decentralized Physical Infrastructure (DePIN) market is transforming the sharing economy by decentralizing physical assets and incentivizing node providers through token-based economics. Central to the DePIN ecosystem are smart contracts that autonomously govern market operations and interactions. Increasingly, AI agents—referred to as \textit{Agentic}~\cite{xi2023rise}—are being deployed within these smart contracts, enabling decision-making processes and infrastructure management without human intervention. By 2024, the market capitalization of DePIN-related projects has surpassed \$10 billion, reflecting significant popularity and investment. As the DePIN market scales, the potential for widespread adoption of these autonomous AI agents offers significant promise for efficiency and innovation in resource allocation and decentralized infrastructure management.
Despite the growing interest and investment in DePIN, academic research on this topic is limited. Existing studies primarily concentrate on the technical aspects of blockchain and decentralized systems~\cite{fan2023towards}, neglecting the distinct characteristics and economic effects of DePIN~\cite{malinova2023tokenomics}. Moreover, despite this growth, there is a notable lack of academic research on DePIN’s foundational aspects and economic impact. The unregulated nature of decentralized markets and the autonomous operation of AI agents within DePIN systems pose risks that could lead to unexpected negative consequences. Without proper analysis and governance, these systems may inadvertently produce outcomes that conflict with societal values, leading to inefficiencies or unintended harms~\cite{acemoglu2018race,acemoglu2021harms}.

\subsection{Unique Challenges in DePIN Research and Development}

The research and development of Decentralized Physical Infrastructure Networks (DePIN) face a number of critical challenges that remain largely unexplored due to the lack of comprehensive models. Currently, no general model exists to capture the full complexity of these decentralized markets and their evolving dynamics, which makes it difficult to ensure that DePIN outcomes align with societal goals and human values. The absence of such a model leaves a significant gap between industry practices and the scientific inquiry needed to rigorously analyze and design DePIN systems. Below, we outline three major challenges:
\begin{enumerate}[label=\alph*., leftmargin=0.5cm]
    \item \textbf{Modeling Dynamic Market Evolution:} Capturing DePIN's market evolution, from inception to large-scale adoption, remains a challenge. Current models either isolate phases or lack detail to account for feedback loops and nonlinear growth. A comprehensive model for both micro and macro levels is needed.
    
    \item \textbf{Identifying Stakeholders and Modeling Interactions:} DePIN markets involve diverse stakeholders with unique incentives. Modeling their interactions is key to understanding market outcomes. A systematic approach is needed to reflect industry practices while providing theoretical abstraction for stakeholder dynamics.
    
    \item \textbf{Macroeconomic Measurement and Value Alignment:} Traditional metrics don't capture the decentralized nature of DePIN markets or their alignment with goals like efficiency, inclusion, and stability. A new framework grounded in real-world practices is needed to assess how market outcomes align with societal values.
\end{enumerate}

\subsection{Our Contributions: The EconAgentic Framework}
\begin{figure}[!htbp]
    \centering
    \includegraphics[width=\linewidth]{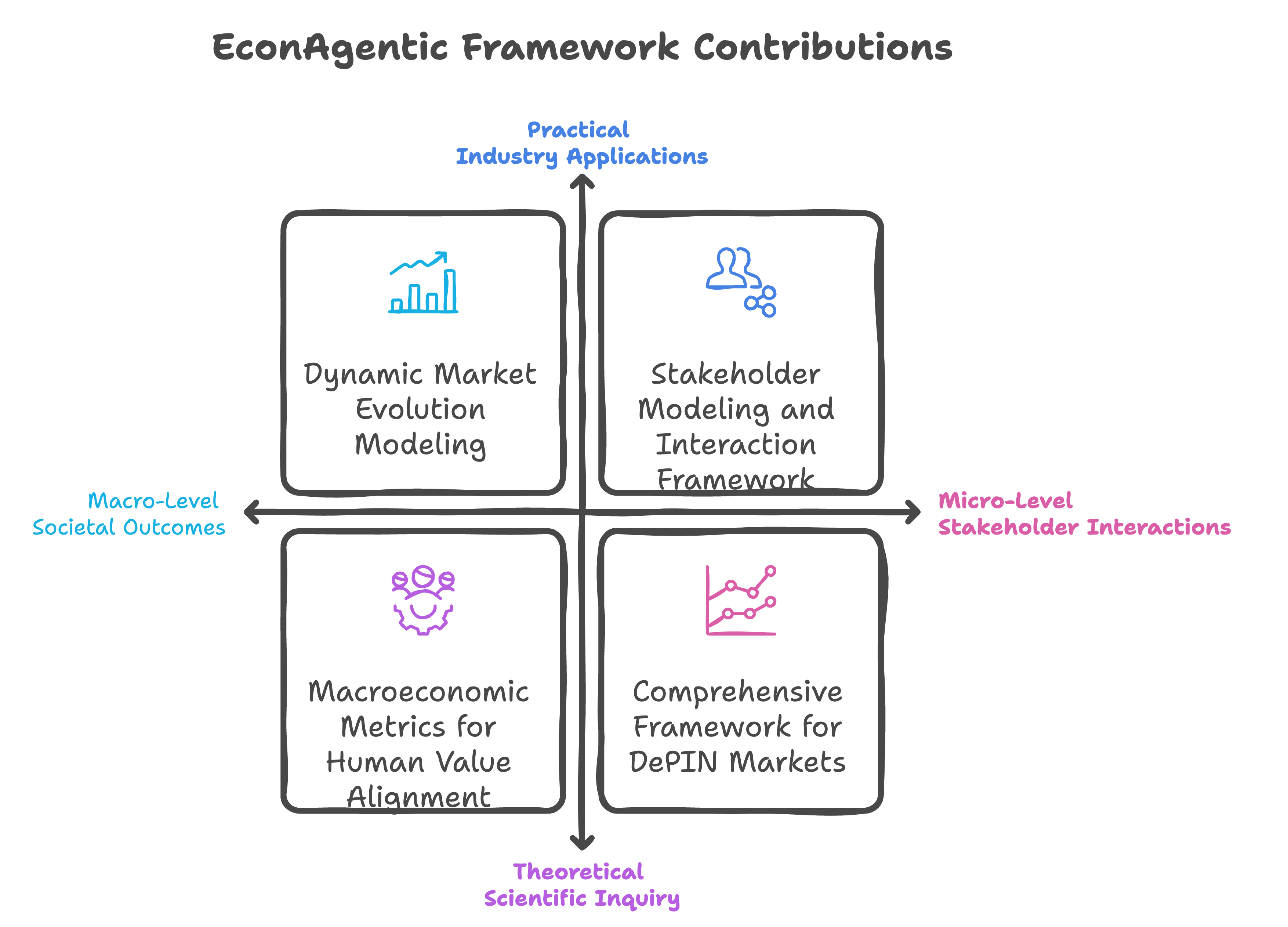}
    \caption{Contributions.}
    \label{fig:fig1}
\end{figure}

To bridge these gaps, we introduce \textit{EconAgentic}, a comprehensive framework designed to empower both academic researchers and industry practitioners to understand, analyze, and design DePIN markets in ways that ensure their outcomes align with human values. Our framework is both grounded in industry practices and capable of abstracting scientific inquiry at the micro and macro levels. As in Figure~\ref{fig:fig1}, the key contributions of EconAgentic are as follows:

\begin{enumerate}[label=\alph*., leftmargin=0.5cm]
    \item \textbf{Dynamic Market Evolution Modeling:} EconAgentic provides a novel methodology for modeling the dynamic evolution of DePIN markets, from their inception to large-scale adoption. Our approach accounts for market fluidity, stakeholder behaviors, technological advances, and regulatory changes. This model is industry-informed but also designed to abstract core principles that govern market growth and feedback loops, allowing for both practical application and scientific exploration.
    
    \item \textbf{Stakeholder Modeling and Interaction Framework:} We propose a systematic approach to identifying and modeling the key stakeholders in the DePIN market and their interactions. EconAgentic allows us to analyze the actions of individual stakeholders and the aggregated impact of their behaviors on the overall market. Our framework balances industry realities with the ability to abstract stakeholder dynamics into a form suitable for scientific inquiry, facilitating better market design and alignment with societal goals.
    
    \item \textbf{Macroeconomic Metrics for Human Value Alignment:} EconAgentic introduces new macroeconomic measurement tools designed specifically for DePIN markets. These tools are grounded in industry practice but abstracted to assess aggregate market outcomes at a high level, ensuring alignment with human values such as fairness, sustainability, and equity. This novel measurement framework enables policymakers and industry practitioners to guide DePIN markets toward socially beneficial outcomes.
\end{enumerate}

In summary, \textit{EconAgentic} provides a much-needed framework for bridging the gap between industry practices and scientific research in the context of DePIN markets, while addressing the unique challenges these markets present. Our framework is designed to enable a rigorous analysis and design of DePIN systems that prioritize alignment with human values at both micro and macro levels.

The rest of the paper is organized as follows: Section~\ref{sec: market} provides an overview of DePIN markets and models their dynamic evolution. Section~\ref{sec:agent} introduces stakeholder dynamics and presents an LLM agent-based framework for modeling interactions. Section~\ref{sec:macro} focuses on the visualization and analysis of macroeconomic indicators in DePIN markets. Section~\ref{sec:future} discusses related work and suggests directions for future research. The appendix details the data, with an emphasis on geospatial heat maps for managing decentralized infrastructure networks.

\section{Background and Modeling of DePIN Markets}
\label{sec: market}

In this section, we elaborate on how DePIN advances over centralized physical infrastructure (CePIN), the taxonomy of DePIN, and the dynamic evaluation of DePIN markets through different growth stages.
\subsection{The DePIN Advances and Taxonomy}
\begin{figure}[!htbp]
    \centering
    \includegraphics[width=\linewidth]{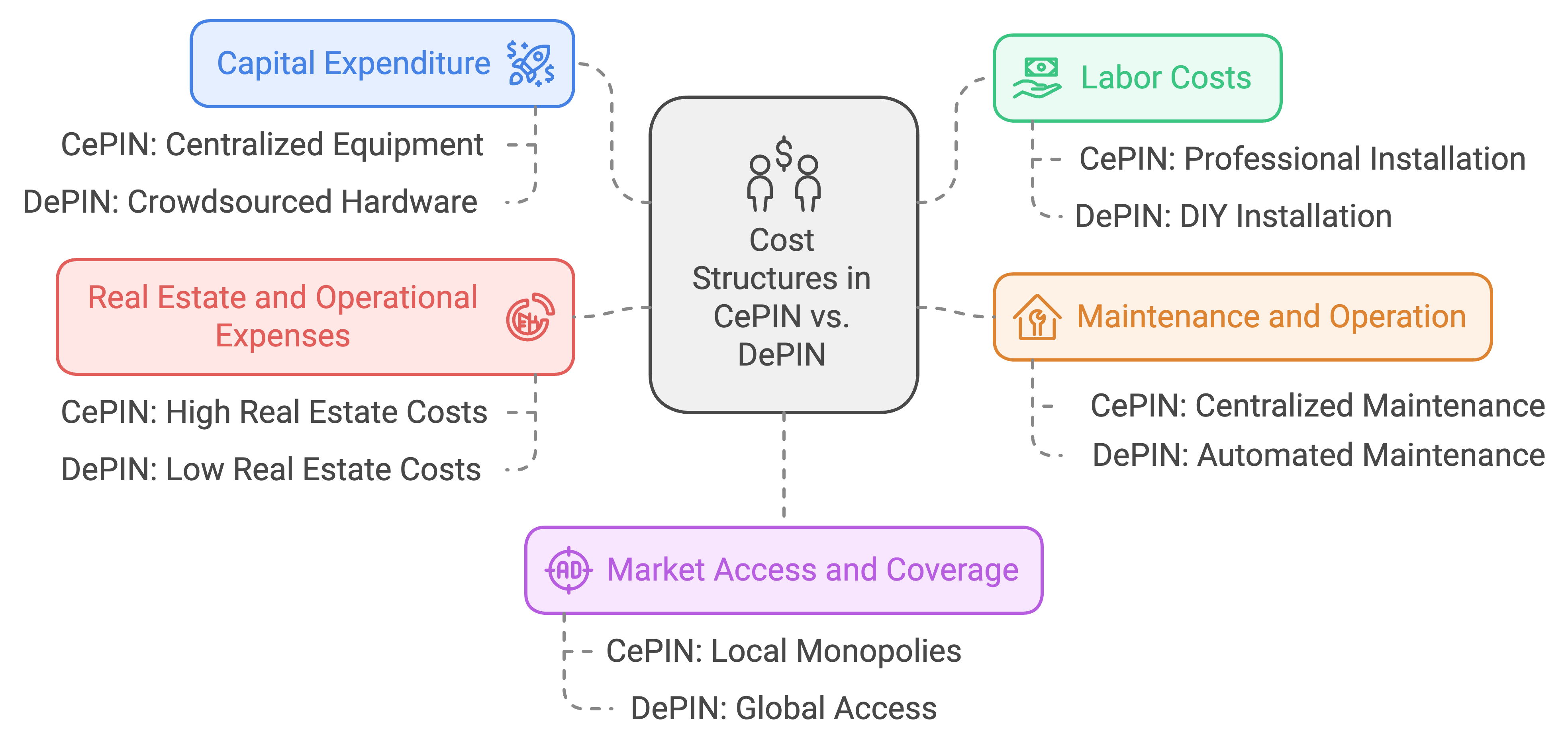}
    \caption{CePIN vs. DePIN.}
    \label{fig:fig2}
\end{figure}

Decentralized Physical Infrastructure Networks (DePIN) offer a transformative approach to managing real-world infrastructure by leveraging blockchain technology and decentralization principles~\cite{iotex2018decentralized}. This shift decentralizes not only service distribution but also the underlying infrastructure layer that coordinates economic activities. DePIN builds upon the success of the sharing economy models, like Uber and Airbnb, by decentralizing the platforms themselves. In DePIN ecosystems, networks of IoT-enabled devices facilitate real-time data exchange and operational decisions, making them smarter and more interconnected~\cite{huckle2016internet,roman2013features}.

As shown in Figure~\ref{fig:fig2}, the transition from centralized physical infrastructure (CePIN) to decentralized models (DePIN) introduces significant advantages across various operational metrics. For instance, DePIN reduces capital expenditure by utilizing crowdsourced, commoditized hardware instead of proprietary equipment. Labor costs are lowered by enabling simplified, DIY installation of hardware, which contrasts with the labor-intensive setups required in traditional networks. Maintenance and operation costs are minimized through decentralized automation and warranty-backed hardware, while real estate and operational expenses are reduced by allowing participants to host hardware on their own premises. Most notably, DePIN offers global market access and flexibility, breaking free from the geographical constraints and monopolies that often limit centralized networks.

This evolution in infrastructure management brings us to the various types of DePIN projects. As outlined in Table~\ref{tab:taxonomy}, DePIN projects can be classified based on the type of physical infrastructure they decentralize. These categories include:

\input{tabs/taxonomy}
\begin{itemize}
    \item \textbf{Server:} Networks that distribute and manage server resources for hosting applications and services across decentralized nodes (e.g., \textbf{Render} on Solana).
    \item \textbf{Wireless:} Networks that enhance wireless communication coverage and resilience using decentralized nodes (e.g., \textbf{Helium} on Solana).
    \item \textbf{Sensor:} Networks that manage IoT and sensor data across decentralized nodes for improved security and integrity (e.g., \textbf{Hivemapper} on Solana).
    \item \textbf{Compute:} Networks that distribute computational resources for scalable data processing (e.g., \textbf{Nosana} on Ethereum).
    \item \textbf{Energy:} Networks that support localized energy management and peer-to-peer trading (e.g., \textbf{Arkreen}).
\end{itemize}

Beyond individual projects, DePIN can be divided into two overarching types: \textbf{Physical Resource Networks (PRN)} and \textbf{Digital Resource Networks (DRN)}~\cite{tintinland2021_DePIN}. PRNs, such as \textit{DIMO}\footnote{\url{https://DePINhub.io/projects/dimo}}, focus on physical assets like vehicles, where participants earn tokens by contributing data. DRNs, like \textit{Akash}\footnote{\url{https://akash.network/}}, allow users to rent out unused digital resources such as cloud computing power.

This taxonomy underscores the versatility of DePIN in managing both physical and digital resources, demonstrating the wide array of applications and the potential for increased efficiency, resilience, and security across various industries. As we analyze the evolution of these projects, it is evident that DePIN’s scalability depends on careful planning across different growth stages.

\subsection{DePIN Market Evolution Dynamics}

\begin{figure}[!htbp]
    \centering
    \includegraphics[width=\linewidth]{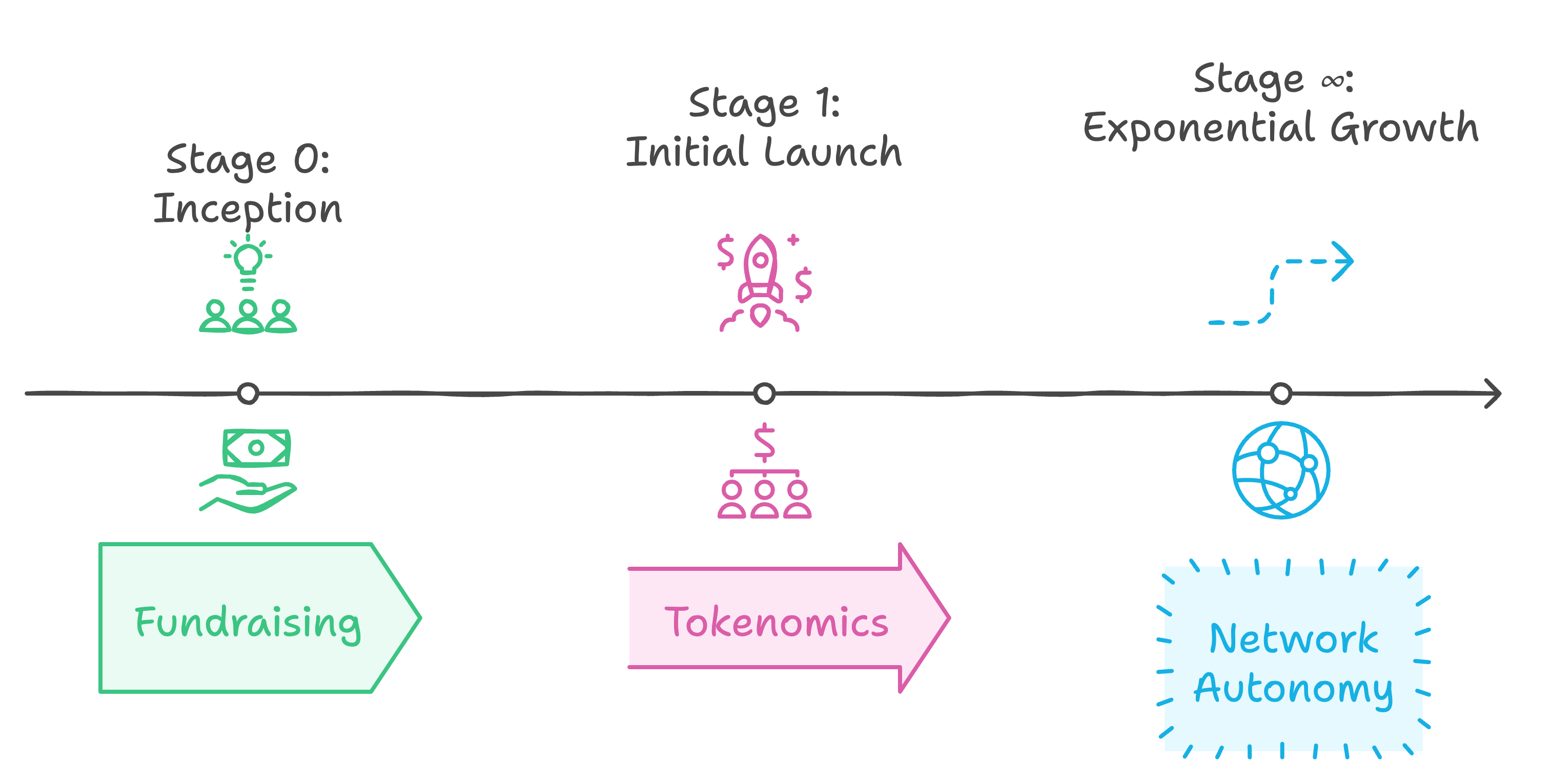}
    \caption{DePIN Market Evolution Dynamics}
    \label{fig:fig3}
\end{figure}

The evolution of a DePIN project follows distinct stages, from its inception to widespread adoption. These stages can be characterized as: \textbf{Inception (Stage 0)}, \textbf{Initial Launch and Scaling (Stage 1)}, and \textbf{Exponential Growth and Widespread Adoption (Stage \(\infty\))} as in Figure~\ref{fig:fig3}. Each stage introduces unique stakeholders, specific actions, and measurable impacts on the project's development and network growth.

\subsubsection{Stage 0: Inception – Conceptualization and Fundraising}

At the inception stage, the core stakeholders include:
\begin{enumerate}[label=\roman*.]
    \item \textbf{Core Team/Developers}: These individuals are responsible for defining the technical vision and building the foundational infrastructure of the DePIN project.
    \item \textbf{Early Investors (VCs, Angels)}: These investors provide the initial capital needed to bootstrap the project.
\end{enumerate}

The key action at this stage is \textbf{fundraising}. For example, Helium raised \$53 million in a Series C funding round led by Andreessen Horowitz in 2021\footnote{\url{https://techcrunch.com/2021/08/10/helium-raises-53m-series-c-led-by-andreessen-horowitz/}}, while Render Network raised \$30 million from Multicoin Capital\footnote{\url{https://cryptoslate.com/render-network-raises-30-million/}}. For simplicity, we assume the project raises \$10 million in a single venture capital round.

Half of the raised funds (\$5 million) are allocated to development and operations, while the other \$5 million is used to deploy the initial network nodes. Setup costs can vary significantly, ranging from \$500 to \$1,000 for a Helium Hotspot miner\footnote{\url{https://www.helium.com/}} to \$5,000 to \$20,000 for Render Network’s GPU-based nodes\footnote{\url{https://rendertoken.com/}}. Assuming a cost of \$100,000 per node, the project deploys 50 nodes with the allocated budget.

From a theoretical standpoint, \textit{Metcalfe’s Law} explains the exponential growth in network value. The value \( V \) of the network is proportional to the square of the number of connected nodes \( n \) (\( V \propto n^2 \))\footnote{\url{https://infoworld.com/article/2973672/metcalfes-law.html}}. With 50 nodes, the DePIN network offers baseline services, establishing the first \textit{network effect}, where the utility of the network grows with the number of connected nodes.

\subsubsection{Stage 1: Initial Launch and Scaling}

As the project moves from inception to launch, additional stakeholders emerge:
\begin{enumerate}[label=\roman*.]
    \item \textbf{Core Team/Developers}: Continue expanding and refining the network infrastructure.
    \item \textbf{Early Adopters and Node Operators}: These participants operate the network’s nodes, contributing to operations while earning rewards.
    \item \textbf{Growth Capitalists (GCs)}: New investors who enter the market by purchasing tokens, supporting liquidity and expansion.
\end{enumerate}

In this phase, \textbf{tokenomics} become critical for sustaining growth. Real-world examples such as Helium and Filecoin typically allocate between 15\% and 25\% of tokens to early investors, with vesting schedules to prevent early sell-offs\footnote{\url{https://docs.filecoin.io/}}.
Node operators also earn revenue through network transaction fees. For example, Helium’s IoT network charges usage-based fees, providing an additional revenue stream\footnote{\url{https://www.helium.com/using-the-network}}. As more nodes join, the network effect strengthens, creating a positive feedback loop, where more participants increase the utility and growth of the network.

\subsubsection{Stage \(\infty\): Exponential Growth and Widespread Adoption}

At this stage, the DePIN network reaches a critical mass, entering a phase of self-sustaining, autonomous growth. The network’s infrastructure and tokenomics, established in earlier stages, allow it to operate with minimal external intervention. The following key dynamics drive this stage:

\begin{enumerate}[label=\roman*.]
    \item \textbf{Growth Capitalists (GCs)}: These investors continue to provide liquidity, which stabilizes the token’s market price and ensures the availability of capital for future expansion.
    \item \textbf{End Users}: The general public and businesses begin using the network at scale, significantly increasing transaction volumes and enhancing the network's utility.
\end{enumerate}

At this stage, \textbf{network autonomy} becomes critical. Governance mechanisms and smart contracts guide the ecosystem’s evolution, while decentralized decision-making ensures the network can adapt and grow without centralized control. This allows the network to continue expanding through a self-sustaining feedback loop.

As predicted by \textit{Metcalfe’s Law}, the network’s value grows quadratically with the number of nodes and users. For example, Helium had expanded to over 800,000 hotspots globally by 2023, providing decentralized IoT services\footnote{\url{https://explorer.helium.com/}}. Similarly, Render connects global users with GPU operators, enabling scalable compute power for tasks such as 3D rendering and machine learning\footnote{\url{https://rendertoken.com/}}.

In summary, \textbf{ecosystem stability} and \textbf{autonomous growth} are key in Stage \(\infty\). The network reaches maturity, where stakeholders, including node operators and token holders, interact seamlessly. The decentralized infrastructure now operates autonomously, driven by the incentives and governance structures set in earlier stages.

This autonomous growth necessitates strong stakeholder interactions and efficient frameworks to manage complex operations within a decentralized environment. In the next section, we delve into \textbf{Stakeholder Dynamics and the LLM Agent-Based Framework}, exploring how artificial intelligence and decentralized agents facilitate these interactions.
 
\section{Stakeholder Dynamics and LLM Agent-Based Framework}
\label{sec:agent}

In this section, we introduce a general agent-based framework for analyzing stakeholder behaviors within Decentralized Physical Infrastructure Networks (DePIN). This framework allows for the simulation of various stakeholder types, including node providers, venture capitalists (VCs), growth capitalists (GCs), and end users. The model evaluates stakeholders' behaviors using either heuristic-based or LLM-based agents and assesses their impact on network growth, token economy dynamics, and market stability. The framework is generalizable across different types of stakeholders, and by comparing simple heuristic-based agents with more complex LLM-based agents, we gain insights into how decision-making strategies influence the overall ecosystem.
\subsection{Token Distribution}
\label{sec:token-distribution}

The total token supply is fixed at \( T_{\text{total}} = 1,000,000,000 \) tokens and is allocated among three key stakeholders: the core team, venture capitalists (VCs), and node providers. The allocation is governed by vesting schedules that ensure long-term engagement and alignment with network growth. The total supply is divided as follows:
\[
T_{\text{total}} = T_{\text{team}} + T_{\text{vc}} + T_{\text{node}}
\]
where:
\[
T_{\text{team}} = 0.20 \times T_{\text{total}}, \quad T_{\text{vc}} = 0.20 \times T_{\text{total}}, \quad T_{\text{node}} = 0.60 \times T_{\text{total}}
\]

The distribution follows specific vesting schedules for each group of stakeholders:
\begin{itemize}
    \item \textbf{Core Team (20\%)}: The core team receives 20\% of the total token supply. The team’s tokens are subject to a 4-year vesting period with a 1-year cliff, meaning:
    \begin{itemize}
        \item No tokens are distributed in the first 11 months (cliff period).
        \item At the end of Month 12 (end of the cliff), 25\% of the allocated tokens are distributed.
        \item The remaining 75\% are vested monthly over the following 36 months.
    \end{itemize}

    \item \textbf{Venture Capitalists (VCs) (20\%)}: VCs receive 20\% of the total supply, following a 2-year vesting period with a 1-year cliff:
    \begin{itemize}
        \item No tokens are distributed in the first 11 months.
        \item At the end of Month 12, 50\% of the VC allocation is distributed.
        \item The remaining 50\% is distributed monthly over the second year (Months 13–24).
    \end{itemize}

    \item \textbf{Node Providers (60\%)}: Node providers are allocated 60\% of the total token supply. Their tokens are distributed over multiple periods, with a halving mechanism every 4 years:
    \begin{itemize}
        \item In the first 4 years (Months 1–48), 50\% of the node provider allocation is distributed equally every month.
        \item In the next 4 years (Months 49–96), 25\% of the total allocation is distributed, with the monthly distribution halved compared to the first period.
        \item So on and so forth.
    \end{itemize}
\end{itemize}

\begin{figure}[!htbp]
    \centering
    \includegraphics[width=0.5\textwidth]{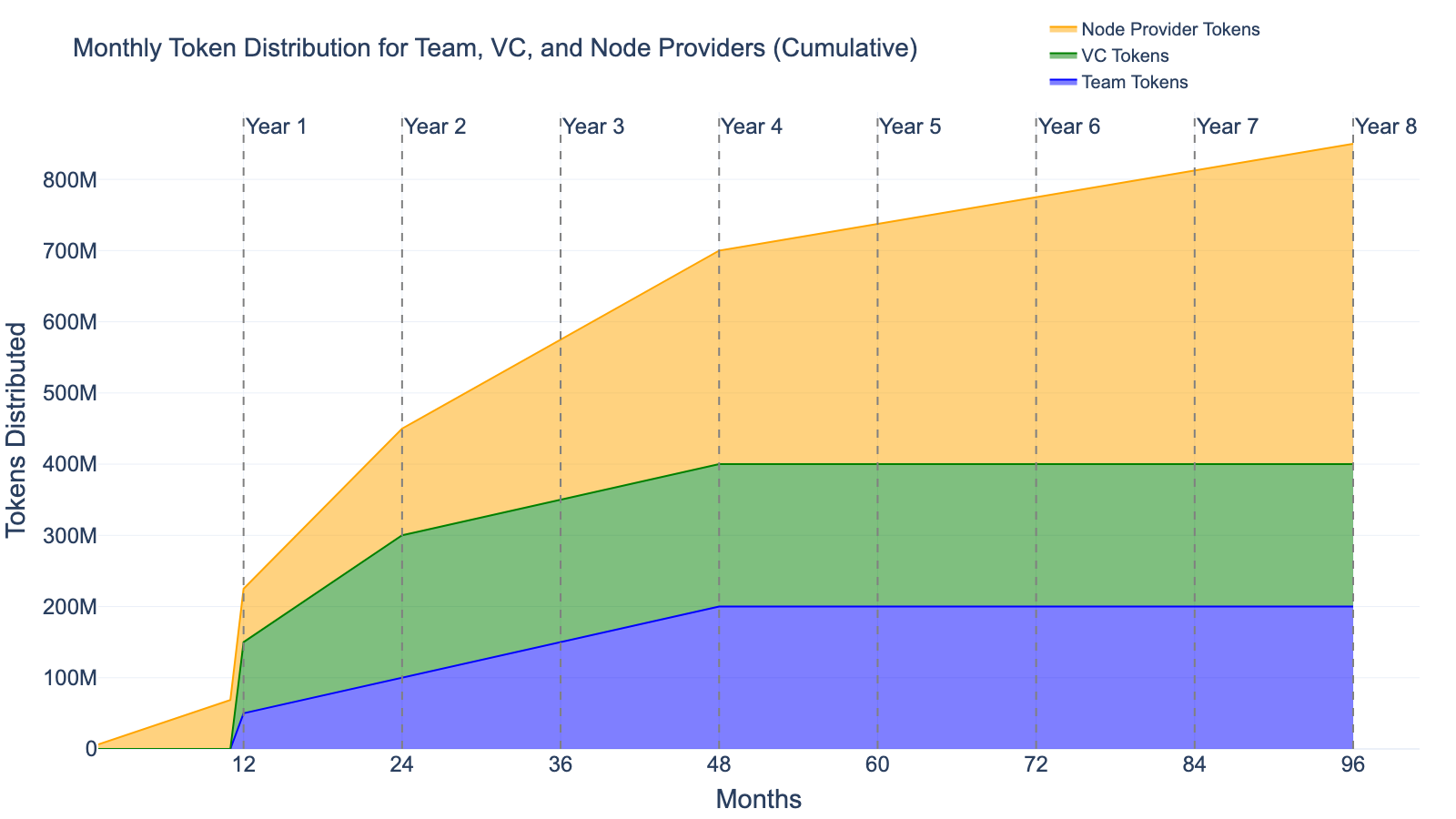}
    \caption{Token distribution schedules for the core team, VCs, and node providers.}
    \label{fig:token-distribution}
\end{figure}

The token vesting schedule ensures that incentives are aligned with long-term participation in the network. The precise distribution of tokens to each stakeholder group is illustrated in Figure~\ref{fig:token-distribution}, showing the vesting periods and halving mechanism for node providers.

\subsection{Simulation Model for Stakeholder Dynamics}
\label{sec:stakeholder-dynamics}

The stakeholder simulation is based on an agent-based model that evaluates the behavior of various participants in the DePIN ecosystem. This model captures interactions between node providers, venture capitalists, growth capitalists, and end users, using both heuristic-based and LLM-based strategies.

\subsubsection{Node Provider Behavior}

The model simulates the entry and exit of node providers based on profitability. The profitability at time \( t \), denoted \( \pi_{\text{node}}(t) \), is calculated as the difference between the global estimated revenue \( R_{\text{global}}(t) \) and the operating cost \( C_{\text{node}} \) for each active node. The formula is given by:
\[
\pi_{\text{node}}(t) = \frac{R_{\text{global}}(t)}{n(t)} - C_{\text{node}}
\]
where \( n(t) \) is the number of active nodes at time \( t \). If a node's profitability exceeds a certain threshold, it remains in the system; otherwise, it exits.

\subsubsection{Global Estimated Revenue}

The total revenue generated by the network, \( R_{\text{global}}(t) \), is influenced by the number of users \( U(t) \), the number of active nodes \( n(t) \), and the token price \( P(t) \). The number of users \( U(t) \) is a function of the number of active nodes, modeled as:
\[
U(t) = 100 \times \sqrt{\frac{n(t) \cdot (n(t) - 1)}{2}}
\]
The global estimated revenue \( R_{\text{global}}(t) \) is then computed as:
\[
R_{\text{global}}(t) = \frac{P(t-1) \cdot T_{\text{node}}(t)}{n(t-1)} + k \cdot U(t)
\]
where \( T_{\text{node}}(t) \) is the number of tokens issued to node providers, and \( k \) is a scaling factor for user-related revenue.

\subsubsection{Growth Capital Dynamics}

Growth capitalists receive endowments \( E_{\text{gc}} \), with lifespans \( L_{\text{gc}} \) drawn from a log-normal distribution. The total endowment at time \( t \), \( E_{\text{total}}(t) \), is the sum of the endowments from active growth capitalists:
\[
E_{\text{total}}(t) = \sum_{\text{gc active at } t} E_{\text{gc}}
\]
Growth capitalists exit after their lifespan \( L_{\text{gc}} \), and ther tokens are sold to the market, influencing the number of tokens on sale.

\subsubsection{Token Price and Market Capitalization}

The token price \( P(t) \) is determined by the ratio of the total growth capital endowment to the tokens on sale at time \( t \):
\[
P(t) = \frac{E_{\text{total}}(t)}{\text{Tokens on Sale}(t)}
\]
Market capitalization \( M(t) \) is then calculated as the product of the token price and the circulating supply:
\[
M(t) = P(t) \times \text{Circulating Supply}(t)
\]
The fully diluted market capitalization, which accounts for the entire token supply, is given by:
\[
M_{\text{diluted}}(t) = P(t) \times T_{\text{total}}
\]

\subsubsection{Simulation Process}

The simulation spans 96 months. At each time step \( t \), the following updates are performed:
\begin{itemize}
    \item Active nodes \( n(t) \) and users \( U(t) \) are updated based on node participation.
    \item Node provider revenue \( R_{\text{global}}(t) \), costs \( C_{\text{node}}(t) \), and profits \( \pi_{\text{node}}(t) \) are calculated.
    \item Growth capitalists enter or exit based on \( L_{\text{gc}} \).
    \item Token price \( P(t) \), market capitalization \( M(t) \), and stability \( \sigma_{\text{log-returns}}(t) \) are adjusted.
\end{itemize}

The simulation outputs key metrics, including token price \( P(t) \), market cap \( M(t) \), and network participation over time, providing insight into system dynamics.

\subsection{Heuristic and LLM-Based Agents}

We compare two types of agents:

\begin{enumerate}
    \item \textbf{Heuristic-Based Agents}: These agents rely on predefined rules to make decisions. For node providers, a node enters the system if the global estimated revenue \(R_{\text{global}}(t)\) exceeds its cost \(C_{\text{node}}\), and exits if the revenue falls below a tolerance threshold.

    \item \textbf{LLM-Based Agents}: These agents use Large Language Models (LLMs) to make decisions based on contextual prompts. LLM-based agents provide more nuanced decision-making by considering broader market conditions and trends.
\end{enumerate}

\subsubsection{Heuristic Node Provider Agent}

The heuristic node provider agent follows a rule-based approach:
\begin{itemize}
    \item \textbf{Node Entry}: A node enters if \(R_{\text{global}}(t) > C_{\text{node}}\).
    \item \textbf{Node Exit}: A node exits if \(R_{\text{global}}(t) < \tau_{\text{node}} \times C_{\text{node}}\), where \( \tau_{\text{node}} \) is the node’s risk tolerance.
\end{itemize}

The pseudocode for the heuristic-based node provider agent is as follows:

\begin{quote}
\boxed{
\begin{aligned}
&\text{For each month } t: \\
&\hspace{0.5cm} \text{a. For each node:} \\
&\hspace{1cm} \text{if } R_{\text{global}}(t) > C_{\text{node}}, \text{ add node.} \\
&\hspace{1cm} \text{if } R_{\text{global}}(t) < \tau_{\text{node}} \times C_{\text{node}}, \text{ remove node.} \\
&\text{Update node count and calculate new revenue.}
\end{aligned}
}
\end{quote}

\subsubsection{LLM-Based Node Provider Agent}

The LLM-based agent provides more context-aware decisions. It generates responses based on natural language prompts to make entry or exit decisions. For example, the decision to enter might depend on a prompt like:

\begin{quote}
    "The global estimated revenue is $R_{\text{global}}$. A node has a cost of $C_{\text{node}}$. Should the node enter the system? Please answer 'yes' or 'no'."
\end{quote}

Similarly, the decision to exit might depend on a prompt such as:

\begin{quote}
    "The global estimated revenue is $R_{\text{global}}$. A node has a cost of $C_{\text{node}}$ and a tolerance of $\tau_{\text{node}}$. Should the node exit the system? Please answer 'yes' or 'no'."
\end{quote}

\section{Visualization and Analysis of Macroeconomic Indicators in DePIN Markets}
\label{sec:macro}

In this section, we analyze key macroeconomic indicators to assess the performance of Decentralized Physical Infrastructure Networks (DePIN) markets. We focus on three critical dimensions: \textbf{Efficiency}, \textbf{Inclusion}, and \textbf{Stability}. These measures offer valuable insights into the economic value, decentralization, and resilience of DePIN markets. Table~\ref{tab:tokens} presents the top 10 DePIN tokens by market capitalization as of May 28, 2024, highlighting key data such as token price, market capitalization, and 24-hour trading volume. The metrics for efficiency, inclusion, and stability are derived from industry-standard practices, providing a comprehensive framework for evaluation.

\input{tabs/tokens}
\subsection{Measures}
\subsubsection{Efficiency}

The \textbf{Efficiency} of a DePIN market is measured by its market capitalization, representing the economic value generated by the network. Mathematically, we define efficiency as the market capitalization \( E_{\text{eff}} \), which is given by:

\[
E_{\text{eff}} = N_{\text{circ}} \times P_{\text{token}}
\]

Where:
\begin{itemize}
    \item \( N_{\text{circ}} \) is the number of circulating tokens.
    \item \( P_{\text{token}} \) is the current token price.
\end{itemize}

Efficiency, represented as \( E_{\text{eff}} \), reflects the total economic value of the network based on the available token data in Table~\ref{tab:tokens}. Higher market capitalization implies a more economically efficient network.

\subsubsection{Inclusion}

The \textbf{Inclusion} of a DePIN market refers to the degree of decentralization and the participation of external node providers. It is critical to evaluate how much of the network's infrastructure is controlled by external participants versus the core team. We represent inclusion using the symbol \( I_{\text{inc}} \), defined as:

\[
I_{\text{inc}} = \frac{N_{\text{ext}}}{N_{\text{total}}} = \frac{N_{\text{total}} - N_{\text{init}}}{N_{\text{total}}}
\]

Where:
\begin{itemize}
    \item \( N_{\text{ext}} \) is the number of nodes operated by external providers.
    \item \( N_{\text{total}} \) is the total number of nodes in the network.
    \item \( N_{\text{init}} \) is the number of nodes initially set up by the core team.
\end{itemize}

The inclusion parameter \( I_{\text{inc}} \) measures the proportion of total nodes operated by external participants. A higher value of \( I_{\text{inc}} \) indicates greater decentralization and participation by external stakeholders. This parameter often requires more technical efforts to measure, as node provider data may not be as publicly accessible as token data.

\subsubsection{Stability}

\textbf{Stability} refers to the volatility of the token price in the DePIN market. Stability is an essential metric for assessing the resilience of the network, as highly volatile token prices may signal instability and risk for investors. We calculate stability \( S_{\text{stab}} \) based on the volatility of token price returns over a specific time period, following the conventional approach of measuring price volatility.

The stability \( S_{\text{stab}} \) is given by the standard deviation of the logarithmic returns of the token price:

\[
S_{\text{stab}} = \sigma_{\text{log-returns}} = \sqrt{\frac{1}{N-1} \sum_{i=1}^{N} \left( \ln\left(\frac{P_i}{P_{i-1}}\right) - \bar{r} \right)^2}
\]

Where:
\begin{itemize}
    \item \( P_i \) is the token price at time \( i \),
    \item \( \ln\left(\frac{P_i}{P_{i-1}}\right) \) represents the logarithmic return from time \( i-1 \) to time \( i \),
    \item \( \bar{r} \) is the average logarithmic return over the observed time period,
    \item \( N \) is the total number of observations.
\end{itemize}

Lower values of \( S_{\text{stab}} \) indicate less price volatility and thus greater stability, which is desirable for long-term network growth and sustainability.

\subsection{Simulation and Results}

In our simulation, we utilized the open-source \textbf{EleutherAI/gpt-neo-125M} language model\footnote{\url{https://huggingface.co/EleutherAI/gpt-neo-125M}} as the decision-making agent within the DePIN economic environment. A key variable in this model is the \textbf{patience} parameter, which governs the number of consecutive signals required for a node to exit the system, thus representing varying levels of risk tolerance. We evaluated LLM strategies with different patience levels and compared them against a heuristic-based benchmark.

\begin{figure*}[!htbp]
    \centering
    \begin{subfigure}[b]{0.32\textwidth}
        \centering
        \includegraphics[width=\textwidth]{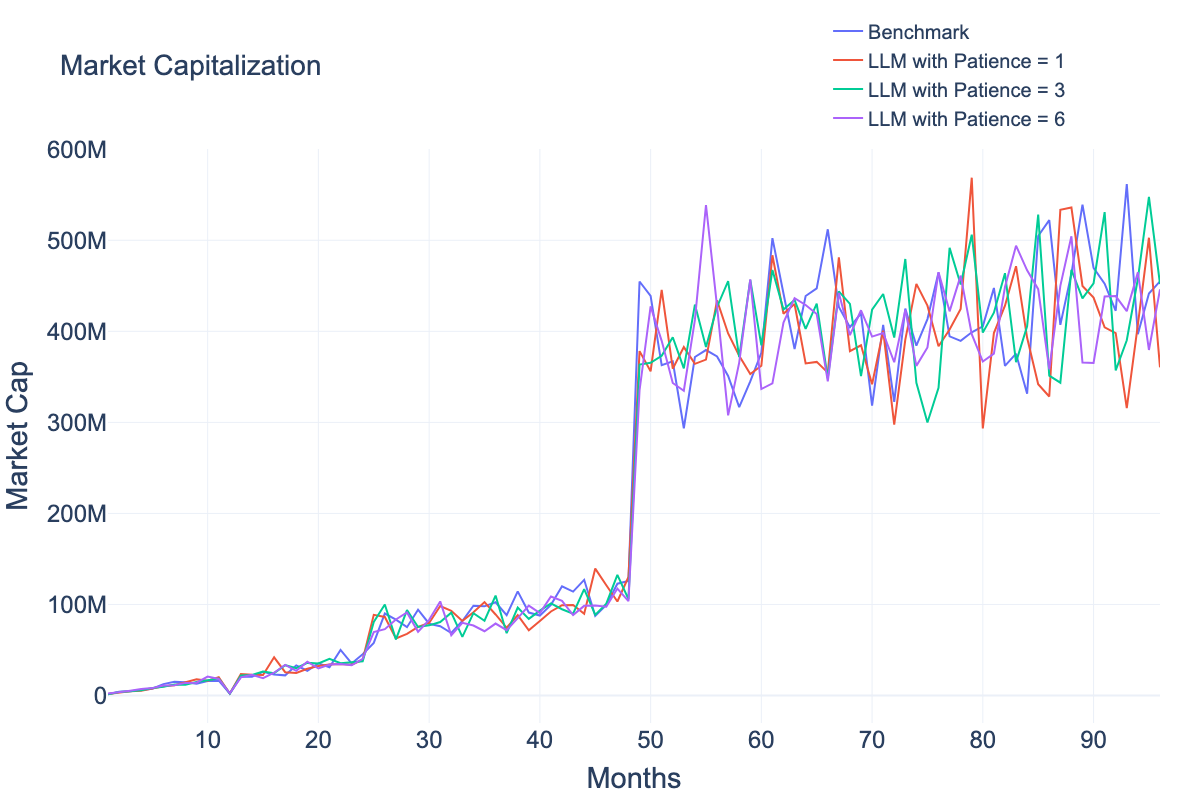}
        \caption{Efficiency: higher values indicate greater efficiency.}
        \label{fig:marketCap}
    \end{subfigure}
    \hfill
    \begin{subfigure}[b]{0.32\textwidth}
        \centering
        \includegraphics[width=\textwidth]{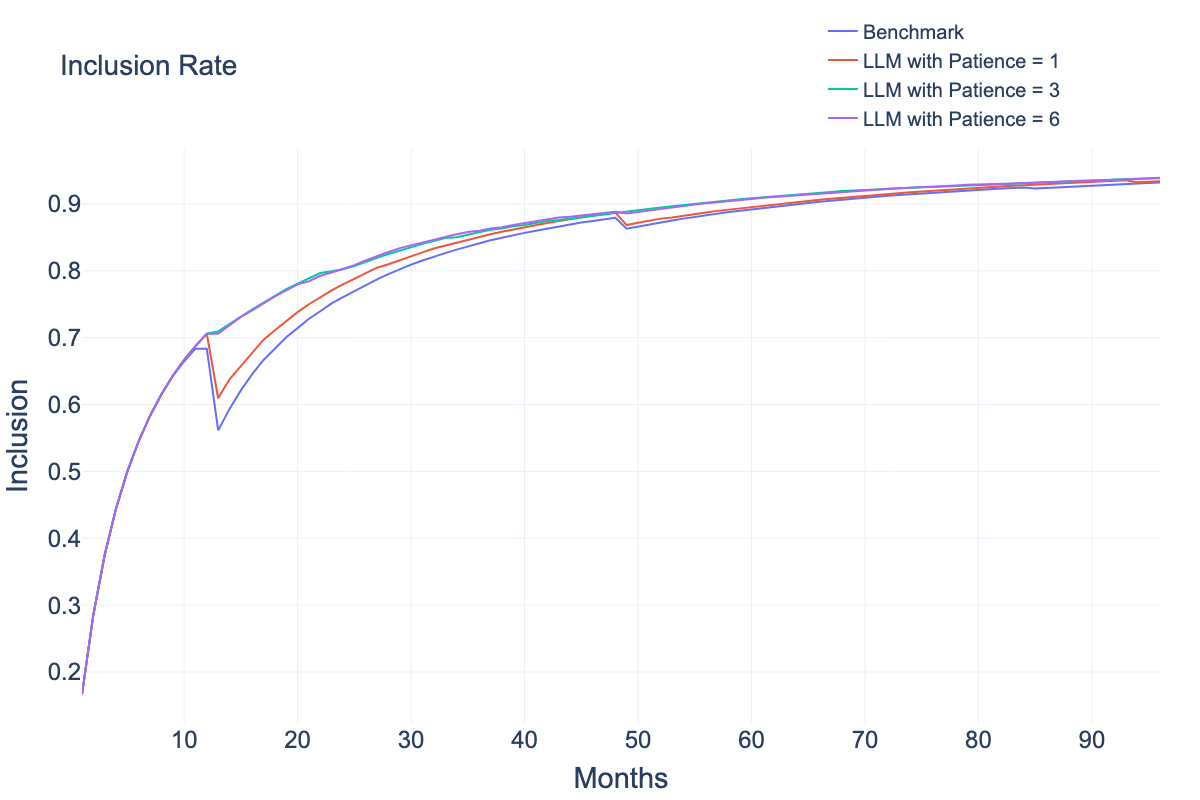}
        \caption{Inclusion: higher values indicate more inclusiveness.}
        \label{fig:inclusion}
    \end{subfigure}
    \hfill
    \begin{subfigure}[b]{0.32\textwidth}
        \centering
        \includegraphics[width=\textwidth]{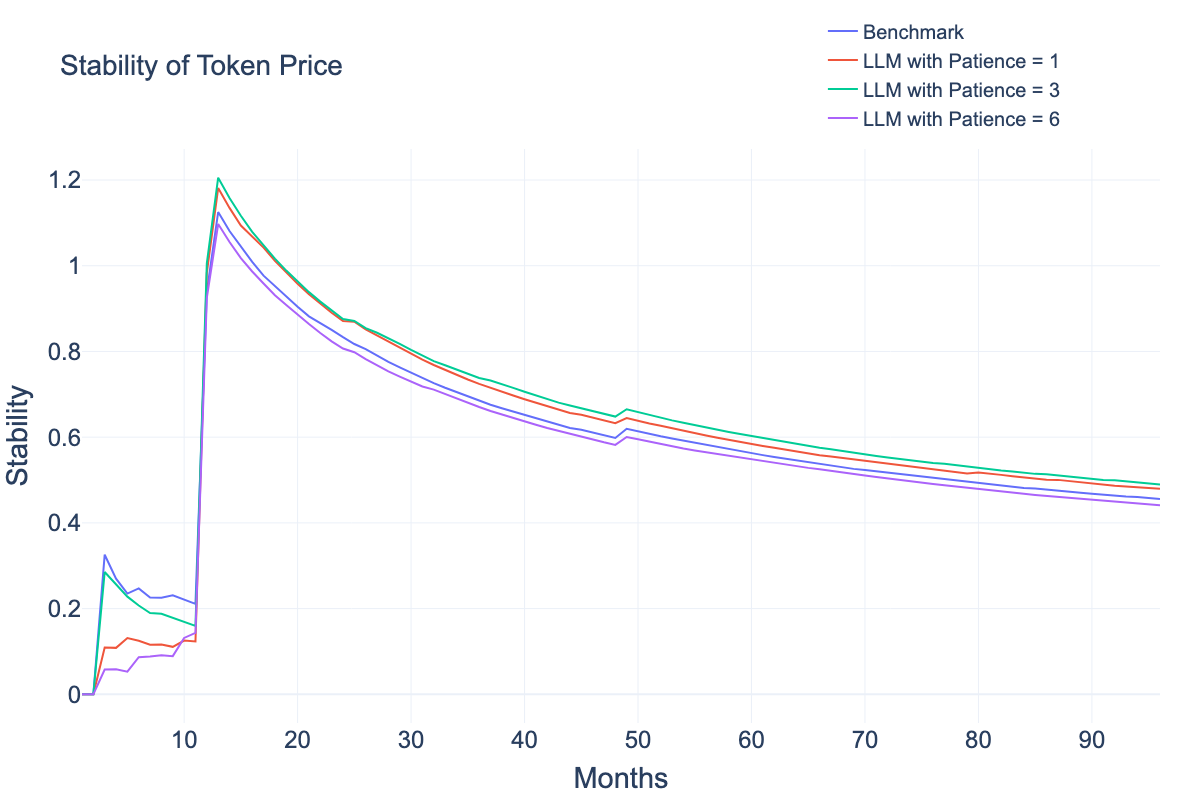}
        \caption{Stability: lower values indicate higher stability.}
        \label{fig:stability}
    \end{subfigure}
    
    \caption{Comparison of Efficiency, Inclusion, and Stability Across Models.}
    \label{fig:measures}
\end{figure*}

Figure~\ref{fig:measures} illustrates the impact of patience on three critical measures: inclusion, stability, and efficiency. \textbf{Inclusion} increases with higher patience, as nodes are retained in the system longer, allowing for more favorable market conditions to emerge. This results in a more inclusive network with fewer premature exits. Similarly, \textbf{stability} improves with greater patience, as nodes become less sensitive to short-term market fluctuations, leading to reduced volatility and smoother system behavior. Although higher patience might theoretically reduce \textbf{efficiency} by keeping non-viable nodes in the system longer, the results show that market capitalization continues to grow across all strategies, indicating that enhanced stability and inclusion do not significantly compromise overall efficiency.

The observed outcomes suggest that LLMs with higher patience are more effective at mitigating short-term volatility, promoting decisions that align with long-term trends. This behavior enhances both inclusion and stability without incurring substantial efficiency losses. Additional visualizations are provided in the appendix.

\section{Related Work and Future Research} \label{sec:future}

This paper makes significant contributions to the study of Decentralized Physical Infrastructure Networks (DePIN) by developing a comprehensive framework that integrates Large Language Model (LLM) agents to evaluate societal goals such as efficiency, inclusion, and stability. Our framework provides a foundation for understanding how stakeholder behaviors influence DePIN market dynamics and establishes benchmarks for assessing the performance of LLM agents in decentralized systems. The following subsections elaborate on our contributions to different strands of the literature, with distinct directions for future research.

\subsection{Token Economy in DePIN}

Chiu et al.\cite{chiu2024DePIN} introduced a foundational framework for DePIN's token economy, focusing on token issuance as rewards and multi-token models that distinguish between value and utility tokens. Their work highlights how staking mechanisms can align short-term behaviors with long-term goals by incentivizing participants to lock tokens into the system. Liu and Zhang\cite{liu2024economics} further analyzed token-based incentives in blockchain governance, demonstrating how staking rewards based on duration can foster long-term participation and stabilize decentralized governance systems.

\textbf{Our Contribution}:
Our contribution lies in setting up a framework that can evaluate the impact of token economies in DePIN markets, particularly in terms of efficiency, inclusion, and stability. While we do not explore specific token design choices like reward structures or staking mechanisms, our framework enables future researchers to assess how these elements influence market outcomes. By focusing on these societal metrics, we provide a foundation for deeper analysis of token economies in decentralized ecosystems.

\textbf{Future Research Directions}:
Future work should utilize this framework to examine how specific token mechanisms—such as reward structures, staking models, and multi-token systems—affect both individual behaviors (microeconomic outcomes) and overall market performance (macroeconomic outcomes). Moreover, there is potential to explore how token designs can be optimized to support broader societal goals, such as equitable resource distribution, environmental sustainability, and ethical governance. Researchers could extend our framework to incorporate fairness and ethics in future evaluations.

\subsection{LLM Agents in Economic Decision-Making}

LLM agents have increasingly been applied to economic and financial decision-making processes. Li et al.\cite{li2024econagent} proposed EconAgent, which uses LLMs to simulate macroeconomic activities and understand complex market dynamics. Other studies, such as ALYMPICS\cite{mao2023alympics} and Ding et al.~\cite{ding2024large}, explored how LLM agents can support strategic decision-making in financial trading by analyzing vast amounts of unstructured data and simulating trading scenarios.

\textbf{Our Contribution}:
We contribute to this line of research by applying LLM agents to DePIN markets, allowing for the simulation and optimization of stakeholder decisions. Our framework enables researchers to analyze how LLM-driven decisions influence key metrics such as efficiency, inclusion, and stability within decentralized, token-based markets. This unique application of LLM agents provides new insights into how AI can support the sustainable development of DePIN systems.

\textbf{Future Research Directions}:
Future research should focus on refining LLM agents to balance short-term economic incentives with long-term network sustainability. Analyzing how LLM agents can assist different stakeholders, such as node operators, venture capitalists, and governance participants, will be essential for improving decision-making processes in DePIN systems. Additionally, extending the application of LLM agents to evaluate broader societal objectives, such as fairness, ethics, and long-term sustainability, represents a promising area for further exploration.

\subsection{Data and Benchmark for LLM}

Our work introduces new benchmarks for evaluating LLM agents in decentralized economic systems, particularly within the context of DePIN markets. These benchmarks allow researchers to assess how LLM agents perform in complex decision-making scenarios, focusing on metrics like efficiency, inclusion, and stability. This effort is inspired by previous benchmarks such as \textbf{BIG-Bench}\cite{srivastava2022beyond}, \textbf{MMLU}\cite{hendrycks2020measuring}, and \textbf{AgentBench}~\cite{liu2023agentbench}, which evaluate LLM capabilities in different domains.

\textbf{Our Contribution}:
We provide a set of benchmarks specifically tailored to DePIN, focusing on societal metrics like efficiency, inclusion, and stability. These benchmarks offer a novel approach to evaluating LLM agents in decentralized, tokenized ecosystems, providing a foundation for future research to assess the sustainability of LLM-driven decisions.

\textbf{Future Research Directions}: Future research should extend these benchmarks to explore how different LLM architectures and training methods impact decision-making in decentralized systems. Additionally, incorporating metrics for fairness, ethics, and long-term societal goals will provide a more comprehensive evaluation framework. Researchers could also examine how LLM agents manage trade-offs between short-term market dynamics and long-term sustainability, further advancing the application of AI in decentralized economies. Future studies could also consider including empirical data for the training process to enhance the realism and applicability of the models. Moreover, more elaboration and cases on decentralized platform data characteristics are provided in the appendix, offering additional insights into the complexities and nuances of such data.

\newpage
\bibliographystyle{ACM-Reference-Format}
\bibliography{DePIN}

\appendix

\section{The Data}
\label{sec: data}
DePIN projects exhibit distinct characteristics that set them apart from traditional infrastructure models. A key feature is the use of geospatial heat maps to visualize the distribution of devices within the network. These heat maps provide a dynamic and intuitive representation of the physical locations and density of decentralized nodes, such as servers, sensors, and wireless devices. By displaying this data spatially, stakeholders can quickly identify areas of high and low device concentration, facilitating more effective resource allocation, maintenance planning, and network optimization. This geospatial approach enhances transparency and monitoring capabilities, supports strategic decision-making, and reveals patterns and trends in device deployment across different geographic regions. Consequently, geospatial heat maps are essential tools in managing and expanding DePIN networks, offering a clear and comprehensive overview of the network's physical footprint. Several DePIN projects provide explorers that include heatmaps of the geospatial distribution of physical infrastructure networks. For instance, Helium provides a hotspot map displaying the distribution of wireless communication nodes, accessible at the \textit{Helium Explorer}\footnote{https://explorer.helium.com/}. 
\begin{figure}[!htbp]
  \centering
  \includegraphics[width=0.5\textwidth]{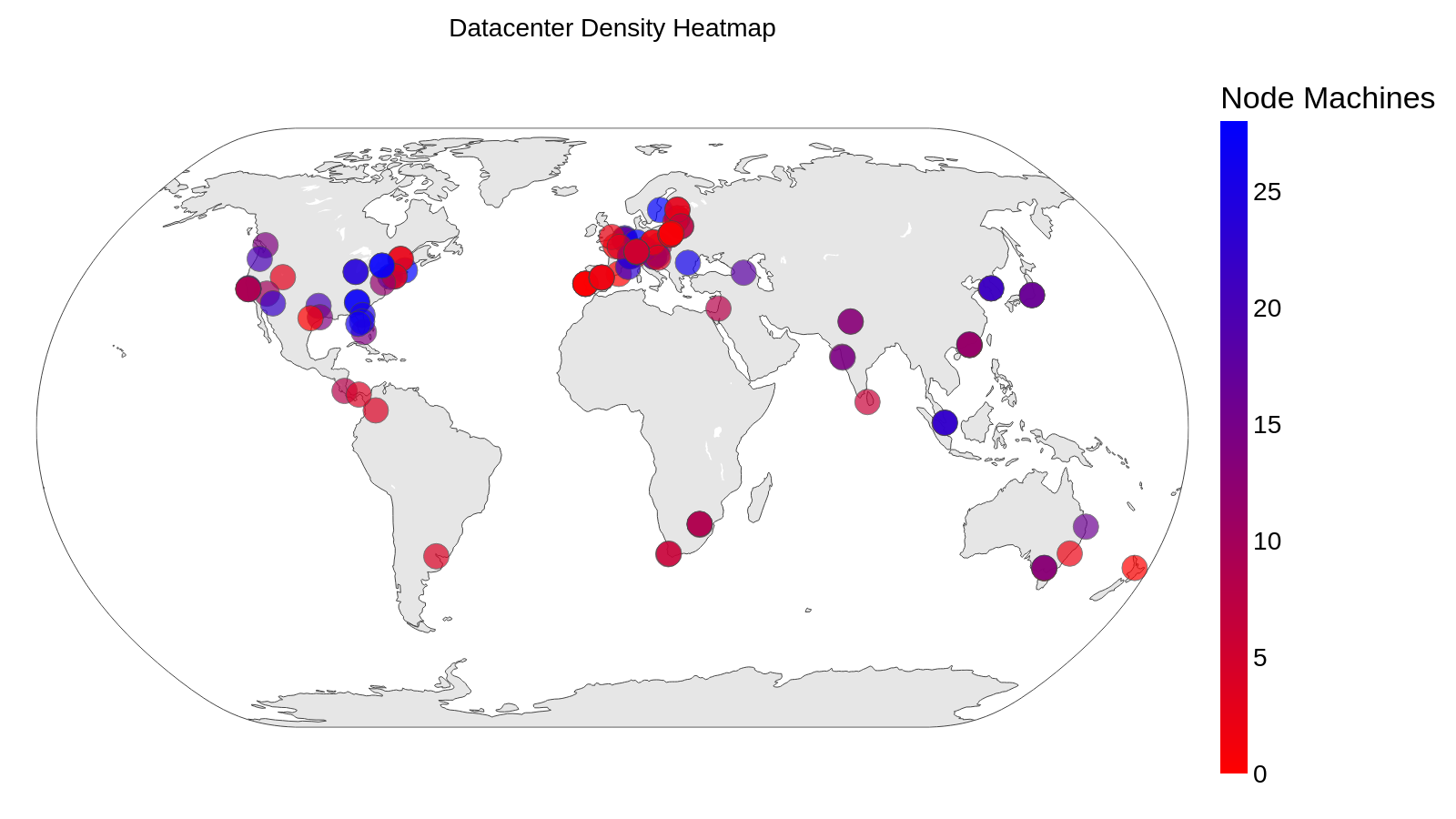}
  \caption{Datacenter Density Heatmap.}
  \label{fig:heatmap}
\end{figure}

\begin{figure}[!htbp]
  \centering
  \includegraphics[width=0.5\textwidth]{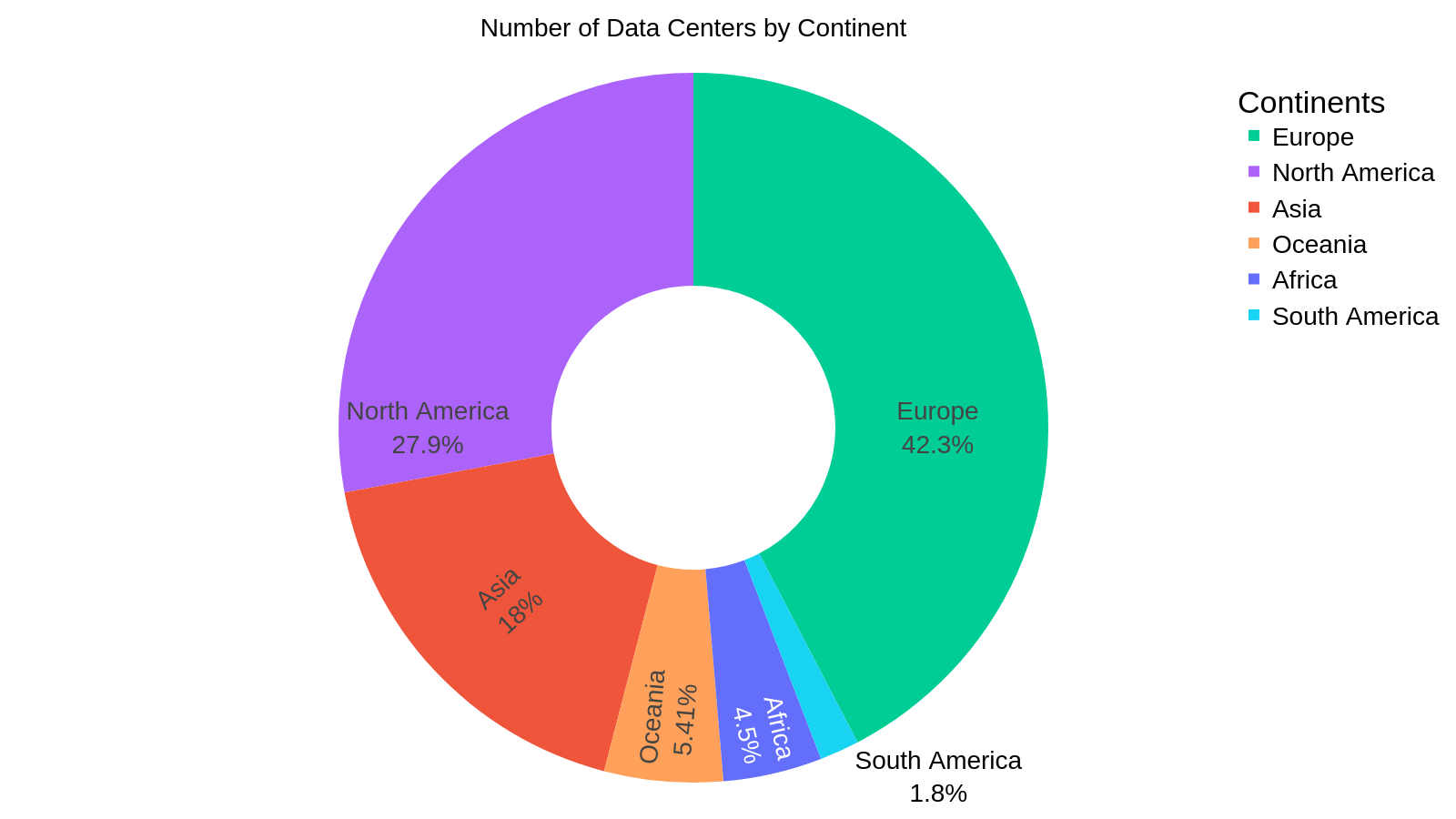}
  \caption{Data Centers by Continent}
  \label{fig:datacenter}
\end{figure}

\begin{figure}[!htbp]
  \centering
  \includegraphics[width=0.5\textwidth]{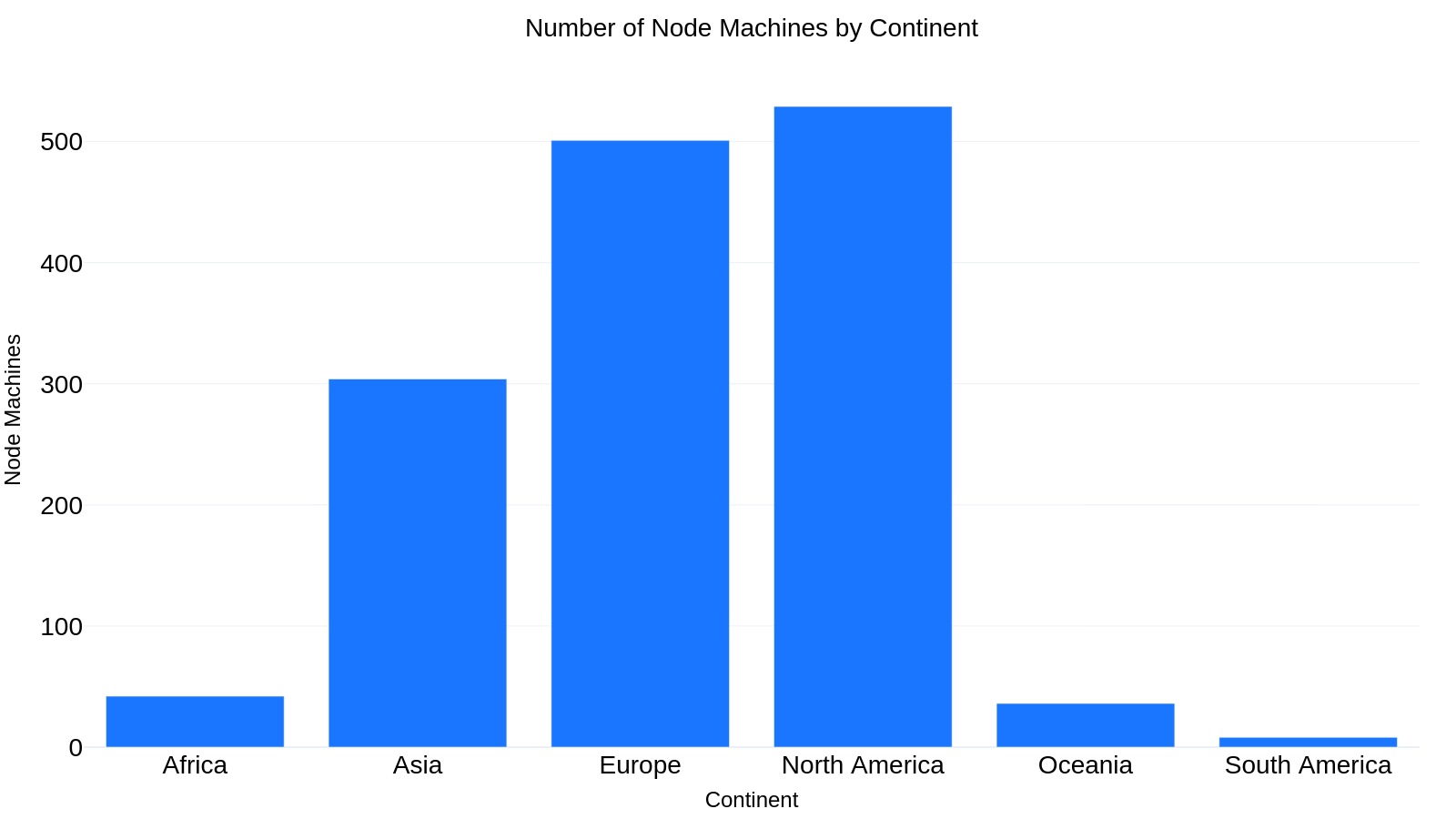}
  \caption{Node Machines by Continent}
  \label{fig:nodes}
\end{figure}

\begin{figure}[!htbp]
  \centering
  \includegraphics[width=0.5\textwidth]{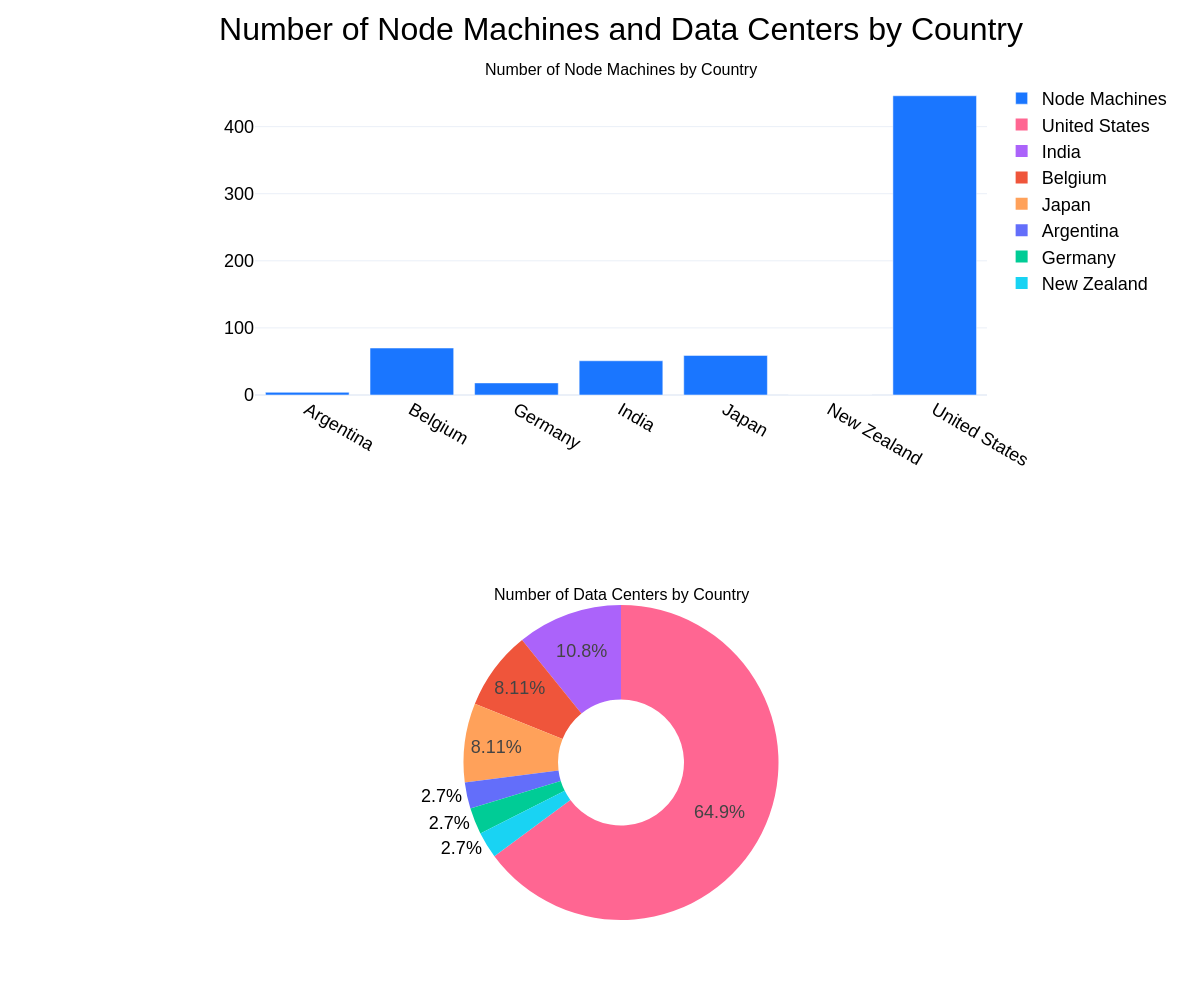}
  \caption{Node Machines and Data Cetners by Countries}
  \label{fig:nodes by countries}
\end{figure}

The Internet Computer protocol, developed by Dfinity Foundation, offers the detailed geospatial data of data center locations, along with the number of active nodes and node providers in each data center, available at the \textit{Internet Computer Dashboard}\footnote{https://dashboard.internetcomputer.org/centers}. Based on this data, we produce visualizations to illustrate the distribution and density of data centers and node machines across different continents.  Figure~\ref{fig:heatmap} shows a heatmap of datacenter density, illustrating the concentration of datacenters around the globe, with colors indicating the number of node machines in each location. Figure~\ref{fig:datacenter} presents the percentage distribution of data centers across different continents, with Europe having the highest number followed by North America. It provides insight into how data centers are geographically distributed among continents. Figure~\ref{fig:nodes} epicts the number of node machines distributed across various continents, with North America and Europe having the highest numbers. It highlights the distribution of computational resources represented by node machines on different continents. Figure~\ref{fig:nodes by countries} shows the node machines and data centers by Coutnries

Another notable platform is DePINscan, which aggregates data from sources like the W3bstream Trusted Metrics API and third-party APIs to provide a comprehensive overview of DePIN projects. Developers can submit their projects for review, and approved projects are featured on the DePINscan explorer. The platform's API supports data integration, allowing the upload of device identifiers and geospatial coordinates (longitude and latitude). This data populates the DePINscan world map, illustrating network device distribution and aiding in understanding deployment patterns. Explore the geospatial distribution of DePIN infrastructure on \textit{DePINscan}\footnote{https://DePINscan.io/map-view}. The major unique contribution of DePIN data lies in extending the application of blockchain technology as a data ledger for digital infrastructure from the purely digital domain to the physical world~\cite{tintinland2021_DePIN}. This transformative approach allows for a secure and transparent recording of geospatial data, bridging the gap between virtual and real-world applications.

However, blockchain technology alone cannot ensure the authenticity of the geospatial data reported onto the blockchain for DePIN projects. Therefore, it is crucial to implement economic mechanisms or tokenomics designs to incentivize all stakeholders, including small businesses and entrepreneurs, to participate. These economic incentives help lower initial costs, enhance efficiency, optimize resource allocation, and reduce waste. Additionally, they improve the integrity of network participants by encouraging honest reporting and active engagement in the ecosystem, thus fostering a more reliable and efficient data ledger system.

In conclusion, the unique characteristics of DePIN data, particularly the use of geospatial heat maps and blockchain technology, offer significant advantages in managing and optimizing decentralized networks. By leveraging these tools, stakeholders can assess and enhance the performance metrics of DePIN networks, ensuring efficient and reliable operation. The integration of economic incentives would further strengthens the network by promoting active and honest participation, ultimately leading to a more robust and effective decentralized infrastructure.

\section{Additional Visualizations}

\begin{figure}[!htbp]
  \centering
  \includegraphics[width=0.5\textwidth]{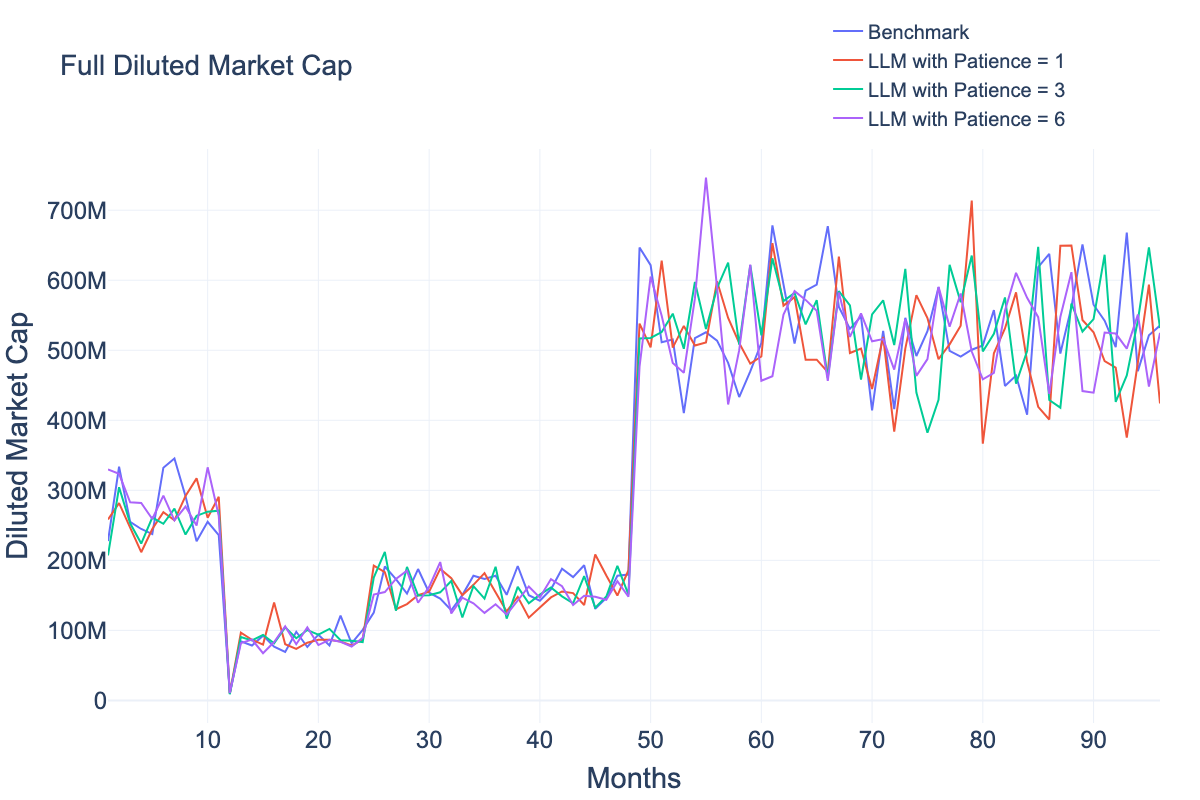}
  \caption{The Diluted Market Cap}
  \label{fig:diluted}
\end{figure}

\begin{figure*}[!htbp]
  \centering
  \includegraphics[width=0.8\textwidth]{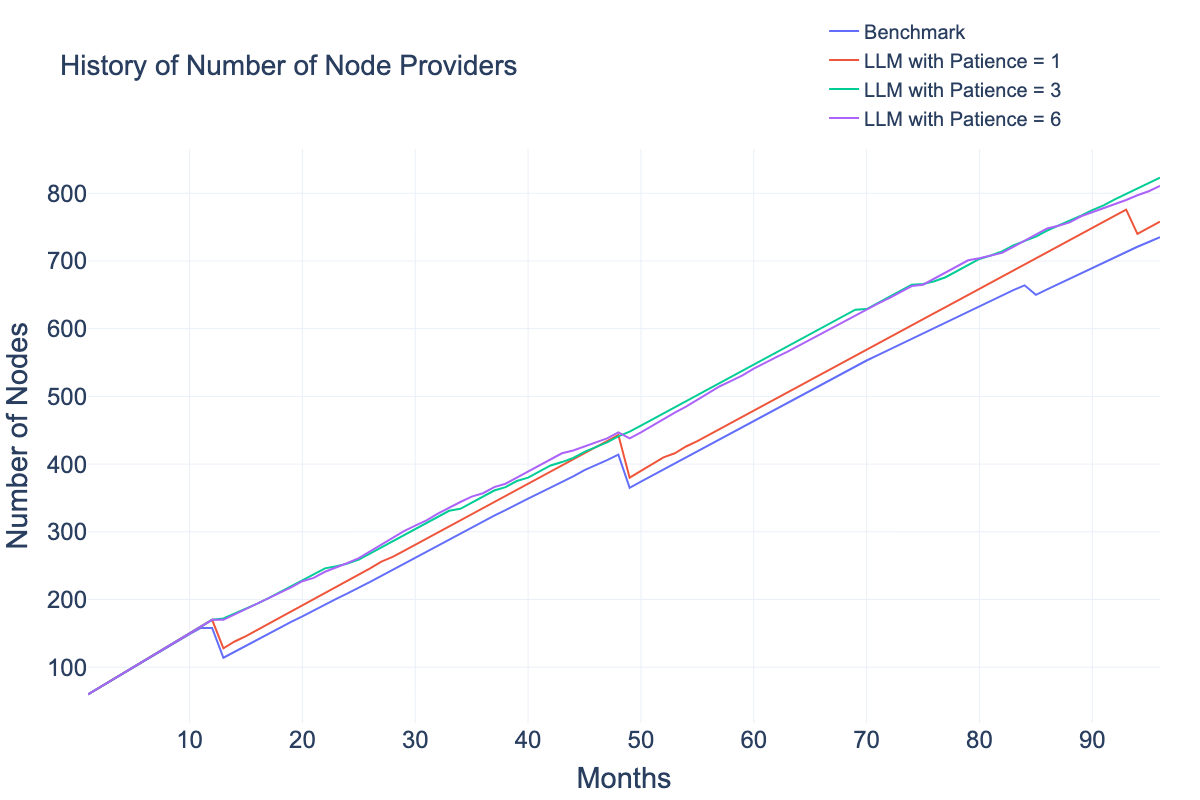}
  \caption{The Number of Nodes}
  \label{fig:num_nodes}
\end{figure*}

\begin{figure*}[!htbp]
  \centering
  \includegraphics[width=0.8\textwidth]{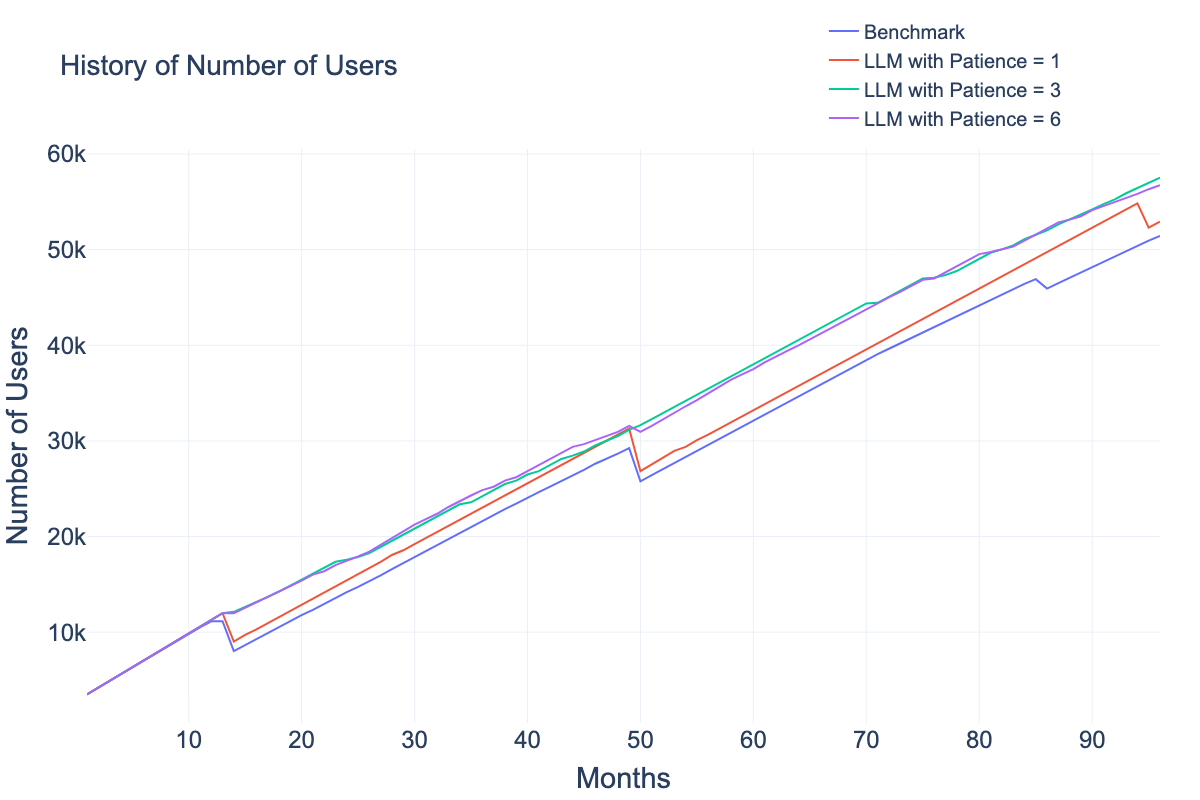}
  \caption{The Number of Users}
  \label{fig:num_users}
\end{figure*}

\end{document}

%% file: tabs/taxonomy.tex
\definecolor{lightgray}{gray}{0.9}
\definecolor{headercolor}{rgb}{0.6, 0.6, 0.6}
\definecolor{rowcolor}{gray}{0.95}

\begin{table}[!htbp]
\caption{Categories of DePIN and their Representative Products}
\centering
\begin{tabular}{|m{1.6cm}|m{3.8cm}|m{1.8cm}|} 
\hline
\rowcolor{headercolor} 
\textbf{\textcolor{white}{Name}} & \textbf{\textcolor{white}{Definition}} & \textbf{\textcolor{white}{Project (Blockchain)}} \\ \hline
\rowcolor{rowcolor}
\textbf{\ding{44}} Server & Networks distribute and manage server resources for hosting applications and services across decentralized nodes. & \href{https://rendernetwork.com/}{Render} (Solana) \\ \hline
\textbf{\ding{115}} Wireless & Decentralized nodes improve coverage and resilience, avoiding central points of failure in communications. & \href{https://www.helium.com/}{Helium} (Solana) \\ \hline
\rowcolor{rowcolor}
\textbf{\ding{46}} Sensor & Networks enhance the management of IoT and sensor data across decentralized nodes for improved security and integrity. & \href{https://hivemapper.com/explorer}{Hivemapper} (Solana) \\ \hline
\textbf{\ding{110}} Compute & Networks distribute computational resources for scalable data processing across multiple decentralized nodes. & \href{https://nosana.io/}{Nosana} (Ethereum) \\ \hline
\rowcolor{rowcolor}
\textbf{\ding{83}} Energy & Networks support localized energy management and peer-to-peer trading, promoting energy independence. & \href{https://arkreen.com/}{Arkreen} \\ \hline
\end{tabular}
\label{tab:taxonomy}
\end{table}

%% file: tabs/tokens.tex
\definecolor{lightgray}{gray}{0.9}
\definecolor{headercolor}{rgb}{0.6, 0.6, 0.6}
\definecolor{rowcolor}{gray}{0.95}

\begin{table}[htbp!]
\centering
\caption{Top 10 DePIN Tokens by Market Cap. Data extracted on May 28, 2024, from CoinMarketCap (\url{https://coinmarketcap.com/view/depin/}).}
\begin{tabular}{|m{2.5cm}|m{1.3cm}|m{1.7cm}|m{1.9cm}|}
\hline
\rowcolor{headercolor}
\textbf{\textcolor{white}{Name}} & \textbf{\textcolor{white}{Price}} & \textbf{\textcolor{white}{Market Cap}} & \textbf{\textcolor{white}{Volume (24h)}} \\ \hline
\rowcolor{rowcolor}
\textbf{Internet Computer (ICP)} & \$12.13 & \$5,631,971,226 & \$92,320,912 \\ \hline
\textbf{Render (RNDR)} & \$10.24 & \$3,980,572,572 & \$236,206,486 \\ \hline
\rowcolor{rowcolor}
\textbf{Filecoin (FIL)} & \$5.94 & \$3,310,041,671 & \$201,176,245 \\ \hline
\textbf{Bittensor (TAO)} & \$416.51 & \$2,850,998,994 & \$30,304,854 \\ \hline
\rowcolor{rowcolor}
\textbf{Arweave (AR)} & \$37.98 & \$2,486,050,282 & \$105,864,760 \\ \hline
\textbf{Theta Network (THETA)} & \$2.27 & \$2,274,255,874 & \$29,805,885 \\ \hline
\rowcolor{rowcolor}
\textbf{Akash Network (AKT)} & \$5.23 & \$1,247,039,844 & \$19,040,451 \\ \hline
\textbf{BitTorrent (BTT)} & \$1.19 \(\times 10^{-6}\) & \$1,152,552,225 & \$28,933,320 \\ \hline
\rowcolor{rowcolor}
\textbf{MultiversX (EGLD)} & \$39.96 & \$1,078,774,444 & \$30,160,556 \\ \hline
\textbf{AIOZ Network (AIOZ)} & \$0.7824 & \$858,001,031 & \$9,057,703 \\ \hline
\end{tabular}
\label{tab:tokens}
\end{table}